\documentstyle[seceq,preprint,epsbox]{jpsj}

\def\runtitle{Third Order Perturbation Analysis of Pairing Symmetry in Two-dimensional Hubbard Model}
\def\runauthor{Hirono {\sc Fukazawa} and Kosaku {\sc Yamada}}

\title
{
Third Order Perturbation Analysis of Pairing Symmetry in Two-dimensional Hubbard Model
}

\author
{ 
Hirono {\sc Fukazawa}\footnote{E-mail: hirono@scphys.kyoto-u.ac.jp}
and Kosaku {\sc Yamada}
}

\inst
{
Department of Physics, Kyoto University, Kyoto 606-8502.
}

\recdate
{
\today
}

\abst
{
We study the superconducting instabilities of singlet and triplet pairing in a two-dimensional Hubbard model on the basis of the third-order perturbation theory (TOPT). We investigate the effect of the vertex correction that is given by TOPT, comparing with the study with only the second-order effective interaction. In our results, a stable $p$-wave pairing state spreads from low to intermediate electron density. A $d_{x^2-y^2}$-wave pairing is dominant for the high density near a half-filling. It is shown that the vertex correction plays an essential role in making the $p$-wave pairing state dominant.
}

\kword
{
Hubbard model, superfluidity, superconductivity, two-dimensional, Fermi liquid, low density, pairing symmetry, perturbation theory
}

\begin{document}
\sloppy
\maketitle
\clearpage

\section{Introduction}
The superconductivity originating from the electron correlation has been investigated in a quasi-two-dimensional system. As a simple model describing the system, a Hubbard model is used in many studies$^{1-8)}$ for a pairing instability. The instability is often studied with a second-order effective interaction with respect to the Coulomb interaction. By these previous studies, it is well known that the dominant state is a $d_{x^2-y^2}$-wave pairing state near a half-filling (the density per spin $n$=0.5). The result is similar to that of the cuprate superconductors. 

The pairing instability has been analyzed also for the electron density lower than the half-filling. Several different consequences are suggested in some analyses. Kondo \cite{rf:3} and Chubukov $et~ al.^{4,5)}$ studied about the pairing instability with the second-order effective interaction. They investigated the Hubbard model based on the weak coupling theory at the zero temperature. 
They indicated the advantage of a $d_{xy}$- and $d_{x^2-y^2}$-wave pairing symmetry. Hlubina \cite{rf:6} reported that a $p$-wave pairing state is possible in the low density at finite temperatures. He indicated that the $p$-, $d_{xy}$- and $g$-wave pairing states are located in the lower density. 

On the other hand, Chubukov\cite{rf:9} considered an influence of the third-order interaction on the $p$-wave pairing at the zero temperature. They studied the problem on the basis of the Fermi gas model. Their results of the two-dimensional system are different from those obtained by the second-order interaction, especially for the $p$-wave pairing. They concluded that the $p$-wave pairing is induced only by the third-order interaction, and the $p$-wave pairing is not encouraged by the second-order interaction possessing a very weak dependence on the wave number. 

Nomura and Yamada\cite{rf:10} made also an analysis for the $p$-wave pairing with the third-order perturbation theory (TOPT).$^{11-14)}$ They investigated a two-dimensional Hubbard model at finite temperatures for the density higher than the half-filling. Their conclusion is similar to the result which Chubukov\cite{rf:9} obtained about the Fermi gas. They concluded that the important factor for $p$-wave pairing is the wave number dependence of the third-order effective interaction, where the third-order interaction for triplet pairing arises from the vertex correction. 

From these results, we consider that the third-order interaction plays an important role for the $p$-wave pairing far from the half-filling in contrast with the analysis based on the second-order interaction. We study the pairing instability at finite temperatures for the density lower than the half-filling. In this paper, we investigate the problem for the two-dimensional Hubbard model on the basis of TOPT. We obtain a phase diagram where the parameters are the hopping integral and the electron density. The phase diagram is compared with Hlubina's\cite{rf:6} and Chubukov $et~al$.'${\rm s}^{4,5)}$ ones. We analyze the role of the third-order interaction together with that of the second-order interaction. Moreover, we actually solve the \'Eliashberg equation to obtain the transition temperature $T_{\rm c}$ of the superconductivity.

Before starting our study, we should discuss the magnetic instability. In quasi-two-dimensional systems, the magnetic instability is considered to be much reduced. Actually there exist superconducting ground states in the quasi-two-dimensional systems such as cuprates, $\kappa$-type BEDT-TTF and Sr$_2$RuO$_4$. Moreover we can expect that the superconducting fluctuations also suppress the magnetic instability. However, it is difficult to determine the actual ground state by using theoretical calculations, since we cannot avoid using different approximations for magnetic and superconducting states, respectively. In this paper, we confine ourselves to the superconducting states in the discussion on the dominant phases.
	
\section{Formulation}
The two-dimensional repulsive Hubbard Hamiltonian for a square lattice is given by
\begin{equation}
{\cal H}=-t_1\sum_{<i,j>,\sigma}c^{\dag}_{i,\sigma}c_{j,\sigma}+t_2\sum_{<i,k>,\sigma}c^{\dag}_{i,\sigma}c_{k,\sigma}+U \sum_{i}n_{i,\uparrow}n_{i,\downarrow},
\end{equation}
where $\sigma$ is the spin index, $<i,j>$ indicates taking summation over the nearest-neighbor sites and $<i,k>$ over the next-nearest-neighbor sites. We obtain the energy dispersion from the non-interacting part in eq. (2.1);
\begin{equation}
E_k=-2t_1(\mbox{cos}k_x+\mbox{cos}k_y)+4t_2\mbox{cos}(k_x)\mbox{cos}(k_y),
\end{equation}
where we take $t_1$=1.0 and $-0.5<t_2<0.5$. By using the above dispersion, we obtain the bare Green's function given by
$G_0(k,\epsilon_n)=\frac{1}{{\rm i}\epsilon_n-(E_k-\mu)}$, where $\epsilon_n=\pi T(2n+1)$ is the Matsubara frequency and $\mu$ is the chemical potential. The particle number (density) per spin $n$ is given by $n=\frac{T}{N}\Sigma_{k,n}G_0(k,\epsilon_n)$.

\begin{figure}
\epsfile{file=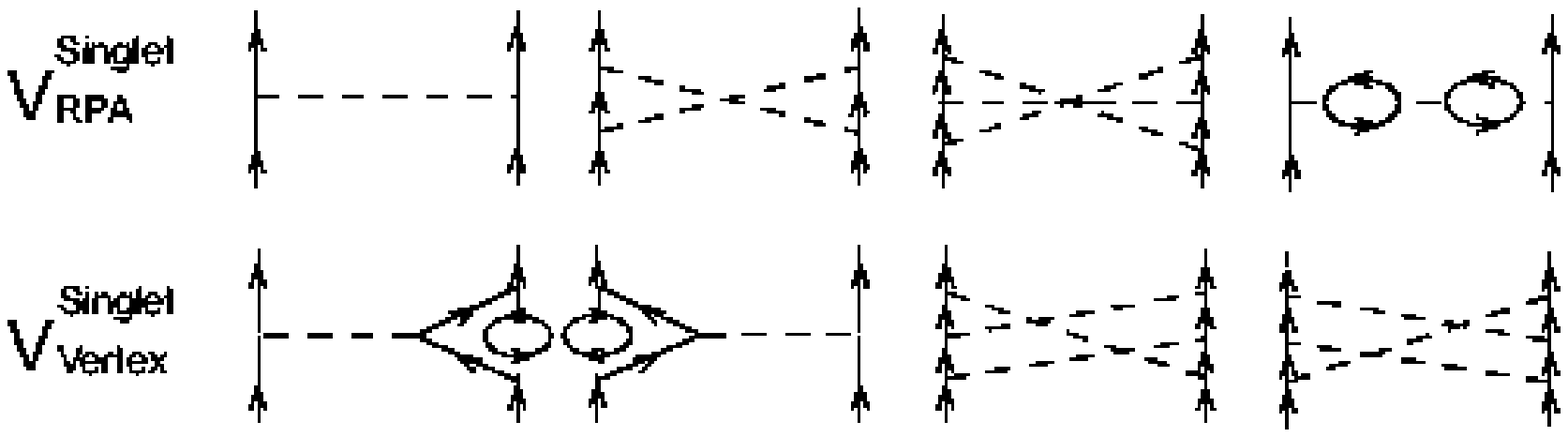,height=4.5cm}
\caption{Diagrams for the effective interaction of singlet pairing within the third-order perturbation with respect to $U$. The solid line is the bare Green's function $G_0$. The broken line is the Coulomb interaction $U$. The broken line of $U$ connects only solid lines possessing opposite spins. The two external lines have the opposite spins. The effective interaction is divided into the RPA-like part and the vertex correction, the latter beginning from the third-order terms.}
\label{fig:1}
\end{figure}

An effective pairing interaction in this study is given by the perturbation expansion up to the third-order term with respect to the on-site Coulomb interaction. We calculate the effective interactions for the singlet and the triplet states, respectively. The effective interaction is divided into two parts,
\begin{equation}
V_{\rm TOPT}(q,k)=V_{\rm RPA}(q,k)+V_{\rm Vertex}(q,k).
\end{equation}
The RPA-like term $V_{\rm RPA}$ includes the term given by the Random Phase Approximation (RPA) and $V_{\rm Vertex}$ is the vertex correction. The RPA-like term reflects the nature of spin fluctuations. The vertex correction term originates from the electron correlation $U$ other than the spin fluctuations. For the singlet pairing, the RPA-part and the vertex correction part are given, respectively, by

\begin{equation}
\hspace{0.0cm}{\it V}_{\rm RPA}^{\rm Singlet}(q,k)=U+U^2\chi_0(q-k)+2U^3\chi_0^2(q-k),
\end{equation}
\begin{equation}
\hspace{0.0cm}V_{\rm Vertex}^{\rm Singlet}(q,k)=2{\frac{T}{N}}U^3[\sum_{k'}G_0(q-k+k')\times(\chi_0(q+k')-\phi_0(q+k'))G_0(k')],
\end{equation}
where $k$ indicates $k \equiv$ (\mbox{\boldmath $k$}, $\omega_n$). The bare susceptibility $\chi_0(q)$  and $\phi_0(q)$ are defined respectively by 
\begin{equation}
\chi_0(q)=-{\frac{T}{N}}\sum_{k}G_0(k)G_0(q + k),~~~~~~~
\end{equation}
\begin{equation}
\phi_0(q)=-{\frac{T}{N}}\sum_{k}G_0(k)G_0(q - k).
\end{equation}
For the singlet pairing, the Coulomb interaction $U$ connects only the electrons which have the opposite spins. The diagrams for the pairing interaction is shown in Fig.~\ref{fig:1}. The two external lines have the opposite spins.

\begin{figure}
\epsfile{file=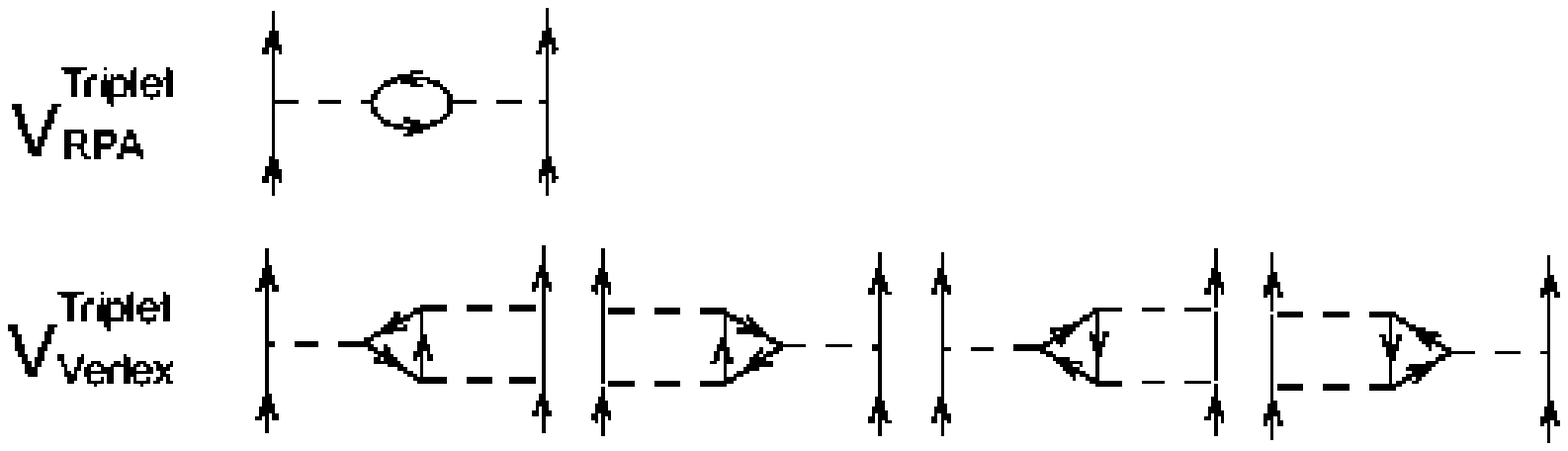,height=4.6cm}
\caption{Diagrams for the effective interaction of the triplet pairing within the third-order perturbation. The two external lines have the parallel spins. The RPA-like part and the vertex correction of the pairing interaction are given by only the second-order and the third-order terms, respectively.}
\label{fig:2}
\end{figure}

For the triplet pairing, the RPA-like term and the vertex correction correspond to the second-order and the third-order terms, respectively.
\begin{equation}
\hspace{0.0cm}V_{\rm RPA}^{\rm Triplet}(q,k)=-U^2\chi_0(q-k), 
\end{equation}
\begin{equation}
\hspace{0.0cm}V_{\rm Vertex}^{\rm Triplet}(q,k)=2{\frac{T}{N}}U^3[\sum_{k'}G_0(q-k+k')\times(\chi_0(q+k')+\phi_0(q+k'))G_0(k')].
\end{equation}
As an effective interaction for the triplet pairing, the diagrams are shown in Fig.~\ref{fig:2}. The two external lines have the parallel spins.

An anomalous self-energy is written by the effective interaction $V(q,k)$ and an anomalous Green function $F(k)$, as $\Sigma_A(q)=-\frac{T}{N}\Sigma_{k}F(k)V(q,k)$. At the transition temperature $T = T_{\rm c}$, the value of the anomalous self-energy $\Sigma_A$ is small and we linearize the \'Eliashberg equation with respect to $F$ and $\Sigma_A$, as $F(k)^{\dagger}=|G_0(k)|^2\Sigma_A(k)^{\dagger}$. From these formulae, we obtain the following equation for the anomalous self-energy;
\begin{equation} 
\lambda\Sigma_A^{\dagger}(q)=-\frac{T}{N}\sum_{k}V(q, k)|G_0(k)|^2\Sigma_A^{\dagger}(k).
\end{equation}
This equation is the linearized \'Eliashberg equation which is an eigenvalue equation with an eigenvalue $\lambda$ and an eigenvector $\Sigma_{A}^{\dagger}$. $V(q,k)$ is given by (2.3), (2.4), (2.5), (2.8) and (2.9). We solve the linearized \'Eliashberg equation on the assumption that $\Sigma_A^{\dagger}$ has the pairing symmetry represented by
\\

\hspace{4.5cm}$p$-wave; \hspace{1.7cm}$\rm sin(k_y)$,

\hspace{4.5cm}$d_{x^2-y^2}$-wave; \hspace{0.8cm}$ \rm cos(k_x)-cos(k_y)$, 

\hspace{4.5cm}$d_{xy}$-wave; \hspace{1.3cm}$\rm sin(k_x)sin(k_y)$,

\hspace{4.5cm}$g$-wave; \hspace{1.7cm}$\rm sin(k_x)sin(k_y)(cos(k_x)-cos(k_y))$.
\\

When the eigenvalue calculated from eq. (2.10) reaches unity, the superconducting state is realized. The most dominant pairing symmetry has the largest value of the eigenvalues among different symmetries. We solve the equation to obtain the dominant state and determine $T_{\rm c}$ at which the eigenvalue $\lambda$ equals unity. 
 
Here, we add a comment on neglecting the normal self-energy $\Sigma_{\rm n}(\vec{q},\omega)$ in our calculation. When we calculate $\Sigma_{\rm n}(\vec{q},\omega)$ with TOPT, $\Sigma_{\rm n}(\vec{q},\omega)$ changes rather drastically for the density far from the half-filling. The change is due to the large asymmetry with respect to electron-hole properties. 
In the low density case where the numbers of electron and hole are very different, the real part of the normal self-energy ${\rm Re}\Sigma_{\rm n}(\vec{q},\omega)$ does not show the smooth Fermi liquid behavior. The wrong behavior of $\Sigma_{\rm n}(\vec{q},\omega)$ is due to the fact that $\Sigma_{\rm n}(\vec{q},\omega)$ is confined to a finite order. The behavior oscillates with respect to the order of $U$, owing to a weak convergence.

To avoid the difficulty due to the weak convergence, we use the following method treating the mass renormalization $z$. The method is discussed in the theory of heavy fermions\cite{rf:15}. By this method, we treat separately the frequency and the momentum dependencies of ${\rm Re}\Sigma_{\rm n}(\vec{q},\omega)$. The frequency dependence of ${\rm Re}\Sigma_{\rm n}(\vec{q},\omega)$ is included by $z$ in the starting Hamiltonian, where $z$ is determined so as to give a correct effective mass enhanced by $z^{-1}$. Then, the momentum dependence is included in the perturbation calculation. 

The detail of the method is the following. The bandwidth $W$ of quasi-particles is renormalized by $z$. The interaction between the quasi-particles $\tilde{\Gamma}$ is renormalized as $z\Gamma z=zU$, where $\Gamma$ is enhanced as $\Gamma=U/z$. As a result, $W$ and $\Gamma$ are renormalized by multiplying $z$. Therefore, the Hamiltonian (2.1) can be rewritten with $t_1$, $t_2$ and $U$ including the renormalization factor $z$. (We don't exhibit the rewritten Hamiltonian explicitly here.) Thus, we can include the frequency dependence of ${\rm Re}\Sigma_{\rm n}(\vec{q},\omega)$ in the starting Hamiltonian. In the procedure for the frequency dependence, we neglect the momentum dependence of ${\rm Re}\Sigma_{\rm n}(\vec{q},\omega)$. 
On the other hand, the momentum dependence of the effective interaction is essential in realizing the anisotropic superconductivity. The momentum dependence is included by solving the \'Eliashberg equation obtained by the perturbation calculation with respect to $U$, as shown in this paper. The separation between frequency and momentum dependencies does not change the essential physics and the procedure of mass renormalization becomes easy.

\section{Numerical Calculation and Results}
We divide the first-Brillouin-zone into 256$\times $256 momentum meshes and take $N_{\rm f}$ = 1024 for Matsubara frequency. To make a reliable calculation, the region of Matubara frequency $\omega_n$ should cover the bandwidth $W$=8. The region is covered with the condition; $|\omega_{\rm n}|<W$. To meet the condition, our calculation is confined to the temperature region $T>0.002$. In addition to the condition, the value of $U$ is confined to $U<W$.

\subsection{Superconductiong phase diagram and $\chi_0(\vec{q},\omega_n = 0)$}
In Fig. 3, we show the phase diagram for the dominant pairing state in the plane of the second-nearest-neighbor hopping integral $t_2$ and the electron density $n$. The regions of parameters are given by $0.075<n<0.4$ and $-0.5<t_2<0.5$. The temperature $T$ equals 0.008 and the Coulomb interaction is $U=$6.0. In Fig. 4, the dependence of the bare susceptibility $\chi_0(\vec{q},\omega_n$ = 0) on $n$ is shown for a quarter first-Brillouin-zone.

For the high density ($n>0.3$) near half-filling in the phase diagram, the $d_{x^2-y^2}$-wave pairing state is dominant due to antiferromagnetic fluctuations which have a peak of susceptibility $\chi_0$ at ($\pi,\pi$) in the momentum space. The situation is the same as that of cuprates corresponding to a nearly half-filled system. The $p$-wave pairing state is realized for the density from $n=0.075$ to $0.303$. The stable $p$-wave pairing state spreads from low to intermediate density ($0.075<n<0.3$).

\subsection{Dependence of $\lambda$ and $T_{\rm c}$ on $n$}
We show the dependence of the eigenvalue $\lambda$ on $n$ in Fig. 5. We solve eq. (2.10) with $T$=0.008, $t_2$=0.25 and $U$=6.4. The region of $n$ is given by $0.03<n<0.375$. The most dominant pairing state has the largest value of $\lambda$ among various symmetries. 

Fig. 5-(a) shows the eigenvalue for eq. (2.10) with $V_{\rm TOPT}$ for all terms which are given by TOPT. The $p$-wave pairing state is dominant over the other states from low to intermediate density ($0.075<n<0.3$). For the high density ($n>0.3$) near the half-filling, the $d_{x^2-y^2}$-wave pairing state is realized. The eigenvalues for the other singlet states (the $d_{xy}$- and the $g$-wave pairing) never reach unity which gives the transition temperature $T_{\rm c}$. 

Fig. 5-(b) shows that $\lambda$ obtained from eq. (2.10) with $V_{\rm RPA}$ given by the RPA-like term, which includes the RPA-like term reflecting the spin fluctuations. The $d_{x^2-y^2}$-wave pairing state is realized for the high density and the $d_{xy}$-wave pairing state is advantageous from intermediate to low density, respectively. In the case of the $p$-wave pairing state, $\lambda$ given by RPA-like term is smaller than one given by TOPT. (The RPA-like term for the triplet includes only the second-order term.) For the singlet pairings, $\lambda$ obtained from the RPA-like term are larger than one of TOPT. From a unity line of $\lambda$ in Fig. 5-(a) for TOPT, we find $T_{\rm c}$=0.008 for the $p$-wave pairing state at $n$=0.075 and for the $d_{x^2-y^2}$-wave pairing at $n$=0.289.

In Fig. 6, we discuss the detail of the eigenvalue calculation and make clear each role of the RPA-like term $V_{\rm RPA}$, the vertex correction $V_{\rm Vertex}$ and TOPT $V_{\rm TOPT}$. The $p$-wave pairing state has the largest eigenvalue for the case possessing only the vertex correction in Fig. 6-(a). (The vertex correction for the triplet includes only the third-order term.) The eigenvalue $\lambda$ for the RPA-like term is the smallest. The eigenvalue $\lambda$ given by TOPT became smaller than one given by the vertex correction. The large eigenvalue obtained from the vertex term is suppressed by the RPA-like term. This is because that a scattering amplitude of $V_{\rm Vertex}^{\rm Triplet}$ has an opposite phase to that of $V_{\rm RPA}^{\rm Triplet}$. About the singlet states in Fig. 6-(b), (c) and (d), the RPA-like term promotes the $d$- and $g$-wave pairing. Thus, these singlet states are encouraged by the spin fluctuations. For example, the antiferromagnetic fluctuations are the main origin for an attractive force of the $d_{x^2-y^2}$-wave pairing. In contrast to the RPA-like term, the vertex correction suppresses the eigenvalue and lowers the transition temperature.

\subsection{Transition temperature $T_{\rm c}$}
We show the superconducting transition temperature $T_{\rm c}$ in Fig. 7. In Fig. 7-(a), we show $T_{\rm c}$ which is calculated with TOPT. The unit of energy is the hopping transfer $t_1$. $T_{\rm c}$ for the $p$-wave pairing state is lower than $T_{\rm c}=0.07$. $T_{\rm c}$ for the $d_{x^2-y^2}$-wave pairing state is higher than $T_{\rm c}=0.03$ for the high density ($n>0.3$) near the half-filling. In Fig. 7-(b), we compare $T_{\rm c}$ which are obtained by TOPT, the RPA-like term and the vertex term, respectively. For the $p$-wave pairing symmetry, $T_{\rm c}$ obtained by only the vertex term is higher than that of TOPT by 0.03. On the other hand, $T_{\rm c}$ given by the RPA-like term for the $d_{x^2-y^2}$-wave pairing symmetry is higher than that of TOPT by 0.2. These fact means that the $p$-wave pairing is induced by the vertex correction and $d_{x^2-y^2}$-wave pairing is induced by the RPA-term.

\subsection{Dependence of $\lambda$ on $U$}
In Fig. 8, 9 and 10, we show $\lambda$ at $U$= 2.0 and 4.0 to study a dependence of $\lambda$ on $U$. The hopping integrals are fixed as $t_2$=0.25 and 0.4. Fig.8 exhibits $\lambda$ obtained by TOPT for the $p$- and $d_{x^2-y^2}$-wave pairing. About the result for $t_2$=0.25, the dominant state at $U=6.4$ is the $p$-wave pairing for about $n<0.3$ and the $d_{x^2-y^2}$-wave pairing in the high density ($n>0.3$) in Fig. 5-(a). The situation does not vary when the value of $U$ changes to 2.0 and 4.0 in Fig. 8-(a). For $t_2$=0.4 (Fig. 8-(b)), the situation does not also vary for the value of $U$. Thus, the value of $U$ does not change the dominant state for the various density in the calculation for TOPT. 

Fig.9 shows $\lambda$ obtained by RPA-like term. At $U$=2.0 and 4.0 for both $t_2$=0.25 and 0.4, the advantageous pairing state is the $d_{xy}$-pairing state from low to intermediate density ($0.075<n<0.3$) and the $d_{x^2-y^2}$-pairing state in the high density. However, the $p$-wave pairing is dominant for the very low density ($n<0.075$) in the case of $t_2$=0.25 for $U$=2.0 and 4.0. For $U$=6.4 in Fig. 5-(b), the $d_{xy}$-pairing is dominant for $n<0.275$. The $p$-wave pairing is not advantageous for the very low density. When $U$ decreases from 4.0 to 2.0, the density for the $p$-wave pairing extends from 0.03 to 0.075. In the very low density for small $U$, we think that the $p$-wave pairing becomes dominant owing to the effect of the ferromagnetic spin fluctuation (paramagnon) or too small eigenvalue for the comparison.

Fig. 10 exhibits the $U$-dependence of the eigenvalue $\lambda$ obtained by including the all (TOPT), vertex and RPA-like terms for the $p$-wave pairing state. Similarly to Fig. 6-(a) for $U$=6.4, the results in Fig. 10 show that the vertex correction mainly promotes the $p$-wave pairing state. The RPA-like effective interaction suppresses $\lambda$ for the $p$-wave pairing. This situation does not change, when $U$ equals 2.0 and 4.0 in the cases of both $t_2$=0.25 and 0.4. 

The results of Figs .8, 9 and 10 indicate that the situation of the dominant pairing state does not change for the value of $U$, except for the very low density in the case where $U$ is smaller than the half of the bandwidth.

\subsection{Wave number dependence of $V$} 
The wave number dependence of the effective interaction $V$ in the momentum space is shown in Fig. 11. The wave number dependence brings an electron scattering near the Fermi surface. The scattering causes the attractive force between electrons on the Fermi surface to realize the pairing state. The point $k=k_F$ on the Fermi surface and $q$ denote the initial state and final state of the scattering, respectively. The lighter color shows a region of the stronger effective interaction which gives the frequent scattering from $k_F$.
 
When the $d_{x^2-y^2}$-wave pairing state is advantageous, the singlet effective interaction is shown in Fig. 11-(a). The RPA-like interaction $V_{\rm RPA}^{\rm Singlet}$ reflects the antiferromagnetic fluctuations and gives the scattering leading to the $d_{x^2-y^2}$-wave pairing state near the Fermi surface. The vertex correction $V_{\rm Vertex}^{\rm Singlet}$ makes the scattering weak for the $d_{x^2-y^2}$-wave pairing. When the $p$-wave pairing state is dominant, the triplet effective interaction is shown in Fig. 11-(b). The vertex correction $V_{\rm Vertex}^{\rm Triplet}$ induces the scattering leading to the $p$-wave pairing near the Fermi surface. The RPA-like interaction $V_{\rm RPA}^{\rm Triplet}$ make the scattering weak. Thus, the RPA-like interaction never gives the strong scattering enough to lead the triplet pairing.

\section{Summary and Conclusions}
We have used the third-order perturbation theory for the two-dimensional Hubbard model and studied the dominant pairing state for the superconductivity. We compare our result with the previous one that includes only the second-order effective interaction.  

In contrast with the case including only the second-order interaction, the region of the $p$-wave pairing state is extended broadly from low to intermediate density($0.075<n<0.303$). Thus, the introduction of the third-order terms gives the important influence on the $p$-wave pairing. The $d_{xy}$- and $g$-wave pairing has been suggested in Hlubina's paper\cite{rf:6} based on only the second-order calculation. However, these singlet states do not appear as the advantageous state in our result by TOPT.
 
We discuss more detailed consequences in the followings. In the triplet state, the vertex correction brings mainly the advantage for the $p$-wave pairing. The vertex correction has the wave number dependence. The wave number dependence causes the scattering near the Fermi surface, which induces the attractive force for the $p$-wave pairing. On the other hand, the RPA-like term suppresses the $p$-wave pairing. The spin fluctuations lower the eigenvalue of the $p$-wave pairing. Thus, the ferromagnetic spin fluctuations do not play a role in realizing the $p$-wave pairing. The situation from low to intermediate density is different from that in the very low density for small $U$. In the very low density, the ferromagnetic spin fluctuations (parramagnon) encourage the $p$-wave pairing. However, the eigenvalue for the paramagnon mechanism is less than that for the vertex correction. 

In the case of the singlet state, the vertex correction discourages the singlet pairing. Thus, the vertex correction lowers the transition temperature. The singlet state becomes dominant by the RPA-like terms reflecting the spin fluctuations, which has been shown by many previous studies. 

Our result might interpret the origin of a quasi-two-dimensional superfluidity of ${ }^3$He$^{8,9,16)}$. We indicate that the $p$-wave pairing becomes dominant owing to the paramagnon in the very low density case. The results agree with the consequence about a two-dimensional superfluidity of ${ }^3$He obtained by previous studies\cite{rf:8}. However, the density corresponding to the realistic two-dimensional ${}^3$He-superfluidity is not the very low density. For example, the density of a realistic three-dimensional ${}^3$He-superfluidity is the quarter-filling near the intermediate density. Therefore, we think that the density of the two-dimensional ${}^3$He-superfluidity is the intermediate density. Thus, the vertex mechanism might give the important effect on the superfluidity rather than that due to the paramagnon. Our results might suggest that the vertex correction plays the important role for the superfluidity of ${}^3$He as well as the ruthenate superconductor. In order to make clear the origin of the  ${}^3$He-superfluidity, it is necessary to consider not only the paramagnon but also the wave number dependence of the vertex correction. 

In addition to this, we mention the effect of the lattice for the intermediate density. The influence of the square lattice is weak in the very low density, because the Fermi surface in the very low density is similar to that of the Fermi gas. On the other hand, the effect of the lattice can not be neglected in the case where the density is not very low. Therefore, it is necessary to compare the results obtained for various lattices on the basis of TOPT. 

Our result might suggest the reconsideration concerning the origin of not only two-dimensional ${ }^3$He superfluidity but also the three-dimensional one\cite{rf:17}. A part of this paper is to be published in proceedings of ISSP meeting.\cite{rf:18}
\section*{Acknowledgment}
The authors thank T. Nomura for the valuable discussion on TOPT and the numerical calculation. The authors are also grateful to Dr. H. Ikeda for the discussion concerning the mass renomarization. H. F thanks Dr. A. Tsuruta for the valuable comments. The numerical computation in this work was carried out at Yukawa Institute Facility.

\clearpage
\clearpage
\clearpage
\begin{figure}
\epsfile{file=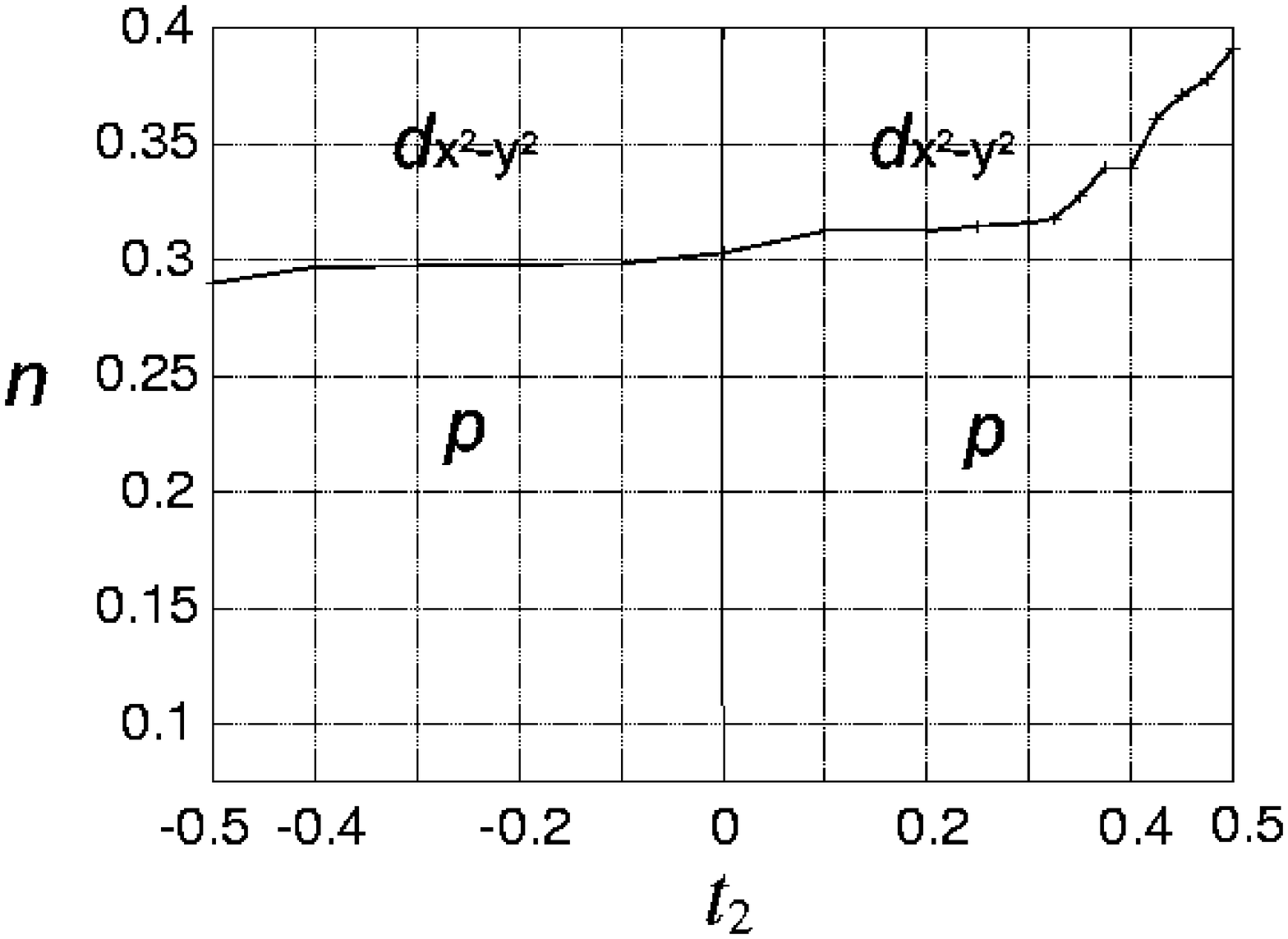,height=10.0cm}
\caption{A phase diagram for the dominant pairing symmetry. The diagram has the parameters of the second-nearest-neighbor hopping integral $t_2$ and the density $n$. $t_2$ is from -0.5 to 0.5 and $n$ is from 0.075 to 0.4. (A half-filling corresponds to the density $n$=0.5.) The temperature $T$ equals 0.008 and the Coulomb interaction $U$ equals 6.0.}
\label{fig:3}
\end{figure}
\clearpage
\begin{figure}
\epsfile{file=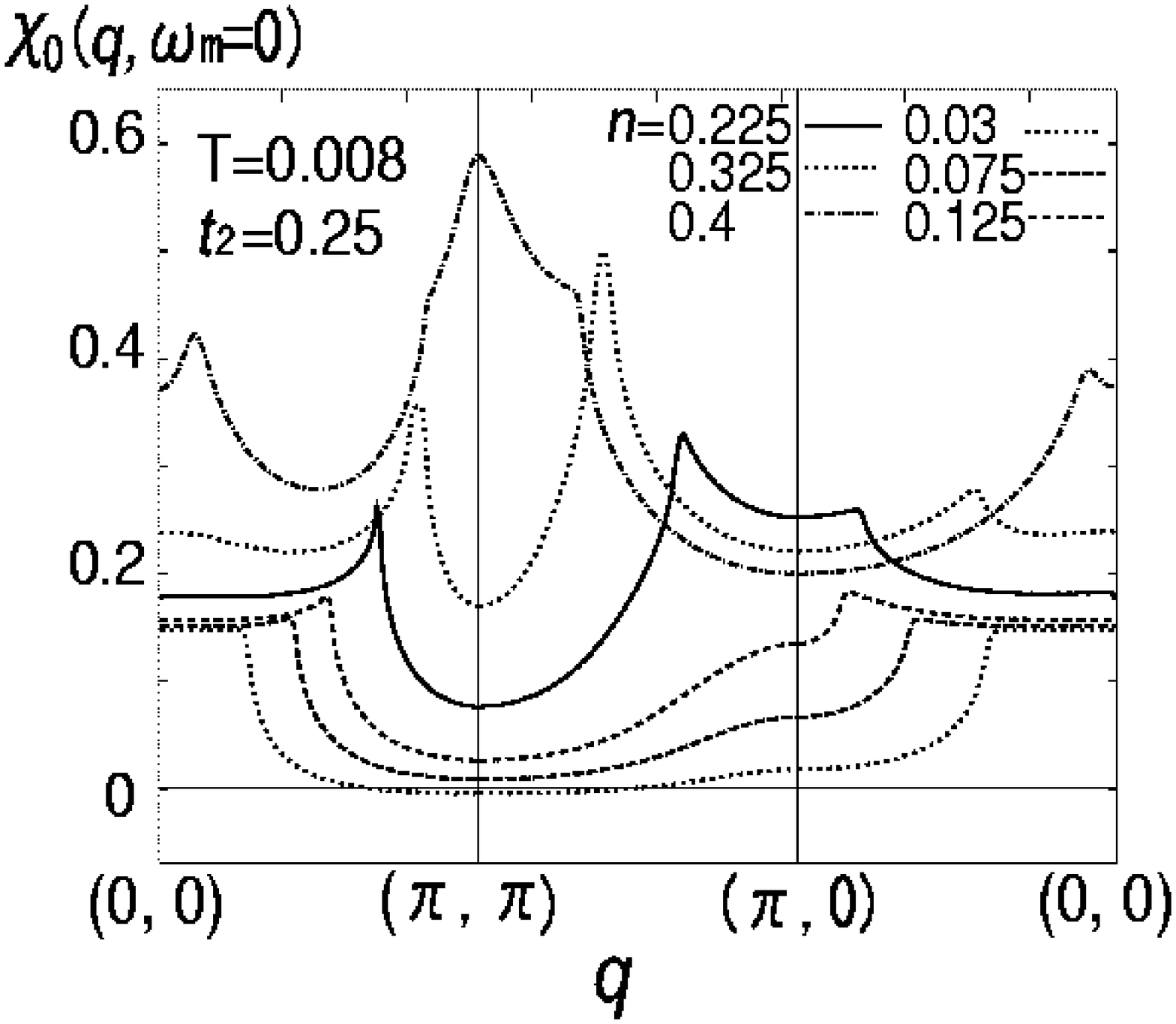,height=12cm}
\caption{A bare susceptibility $\chi_0(\vec{q}, \omega_n=0$). The dependence of the susceptibility $\chi_0(\vec{q}, \omega_n=0$) on the density $n$ in a quarter first-Brillouin-zone. The parameter sets are similar to those of Fig. 1.}
\label{fig:4}
\end{figure}	
\begin{figure}
\epsfile{file=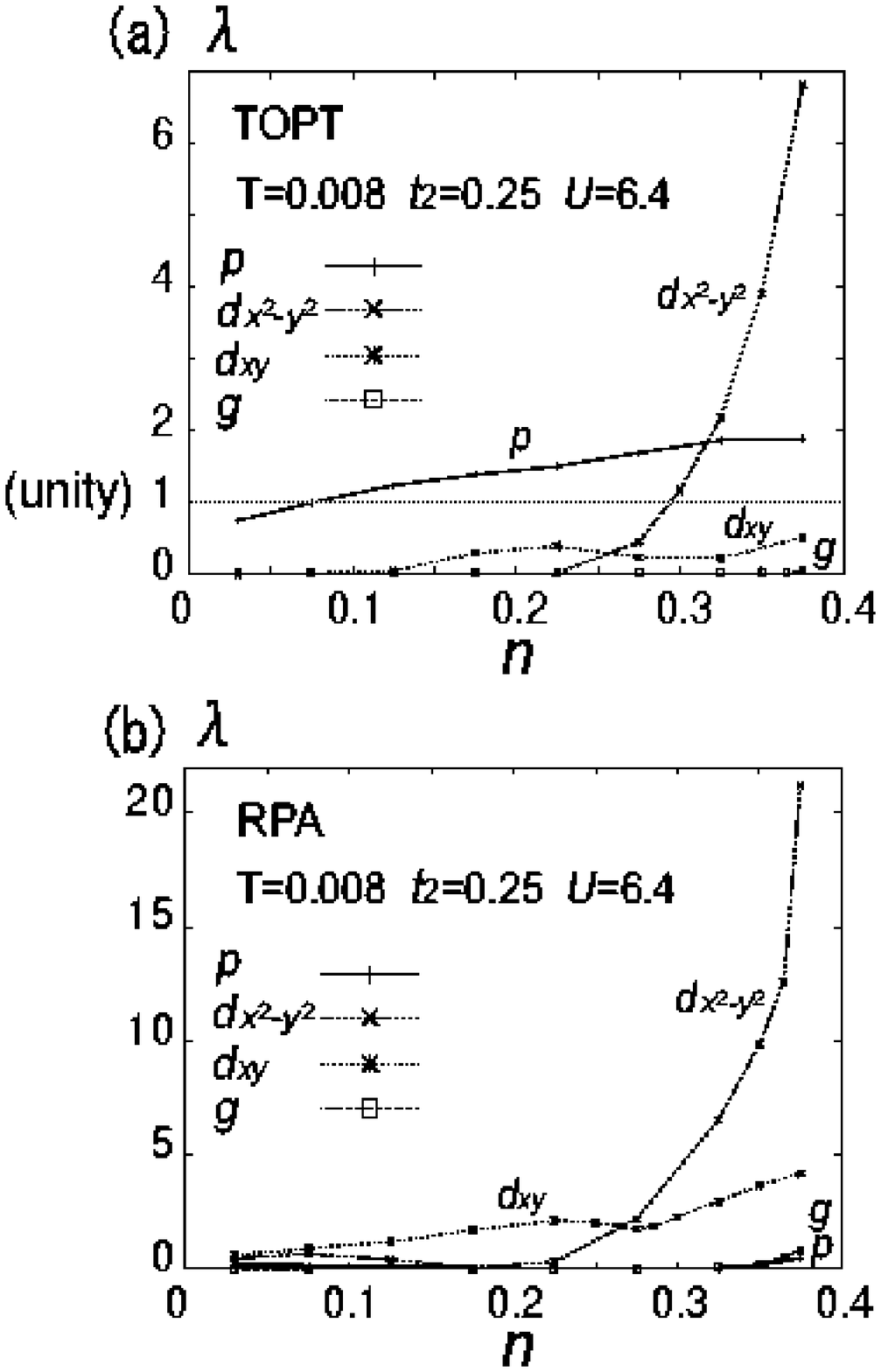,height=20.0cm}
\caption{The dependence of the eigenvalue $\lambda$ on the density $n$. (a) is calculated with all the terms given by the third-order perturbation (TOPT). (b) is obtained from only the RPA-like term. The parameters are the temperature $T$=0.008, the hopping integral $t_2$=0.25, and the Coulomb interaction $U$=6.4. The region of the density $n$ is from 0.03 to 0.375.}
\label{fig:5}
\end{figure}
\begin{figure}
\epsfile{file=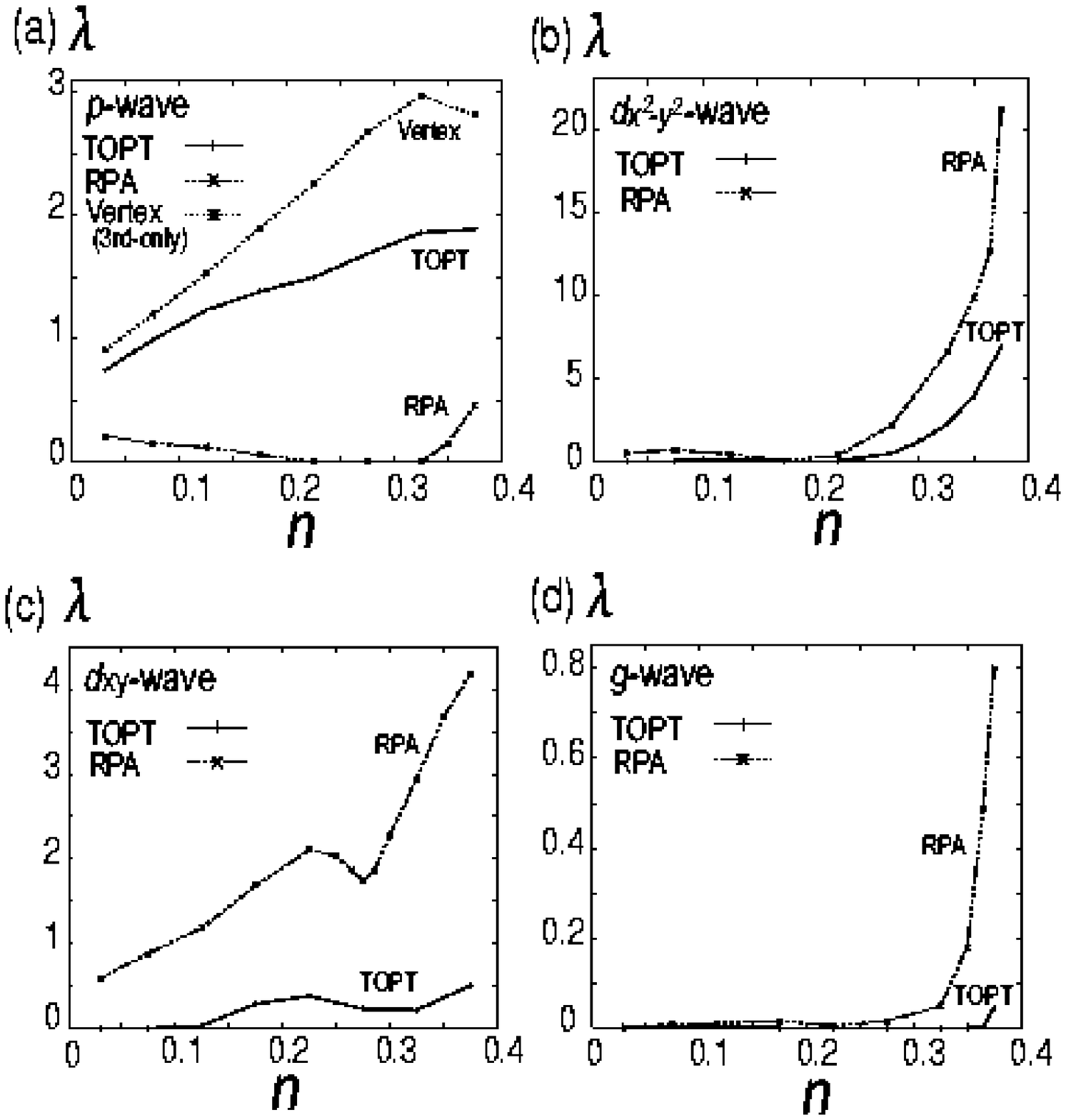,height=18cm}
\caption{The dependence of the eigenvalue $\lambda$ on the density $n$ for (a) $p$-, (b) $d_{x^2-y^2}$-, (c) $d_{xy}$- and (d) $g$-wave pairing state. The parameter sets are similar to those of Fig. 5.}
\label{fig:6}
\end{figure}
\begin{figure}
\epsfile{file=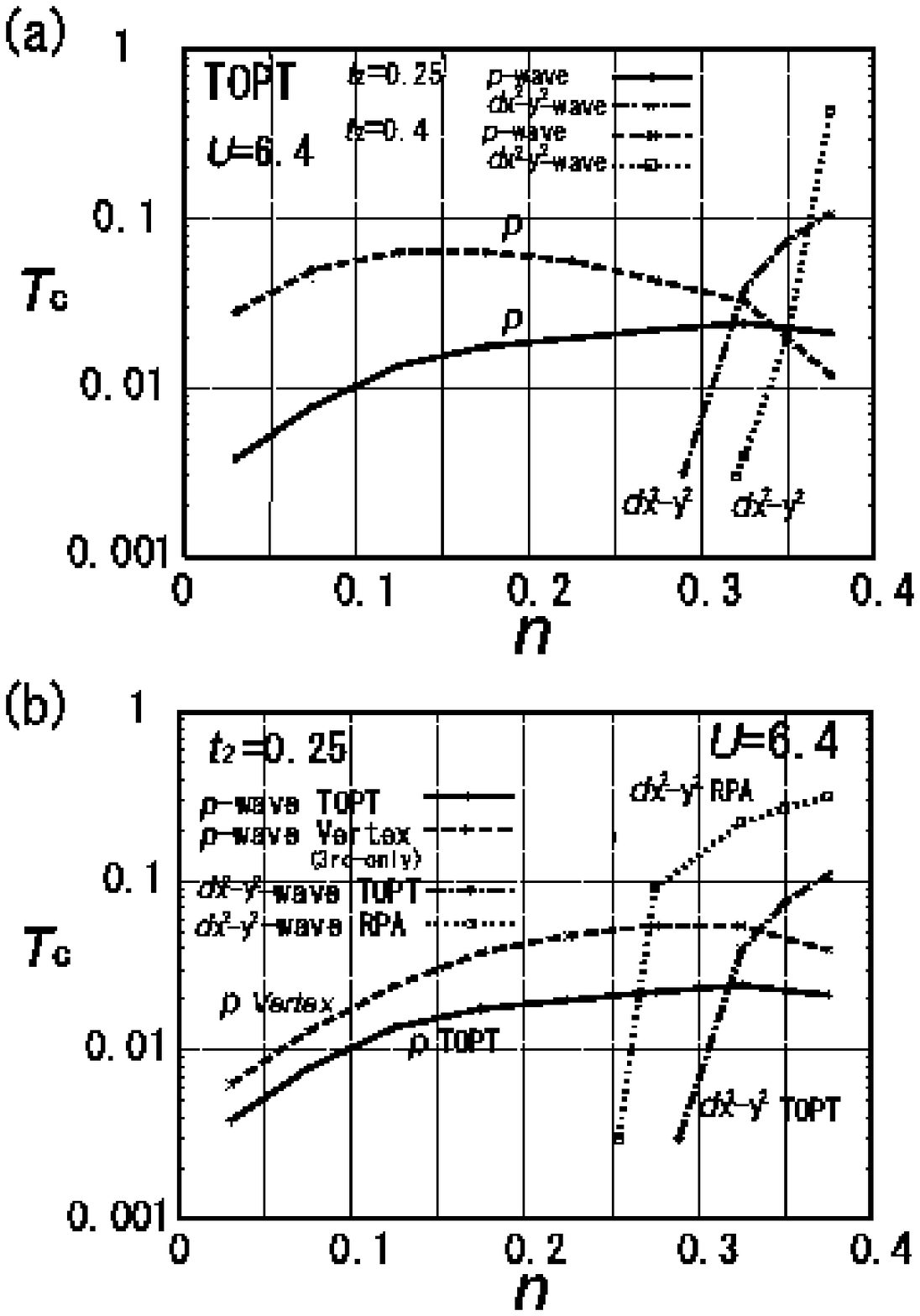,height=18.0cm}
\caption{The superconducting transition temperature $T_{\rm c}$. (a) $T_{\rm c}$ which is calculated with all terms (TOPT) given by the third-order perturbation theory. (b) The comparison of $T_{\rm c}$ obtained with all terms (TOPT), the RPA-like term (RPA) and the vertex term (Vertex). The unit of energy is the hopping transfer integral $t_1$.}
\label{fig:7}
\end{figure}
\clearpage
\begin{figure}
\epsfile{file=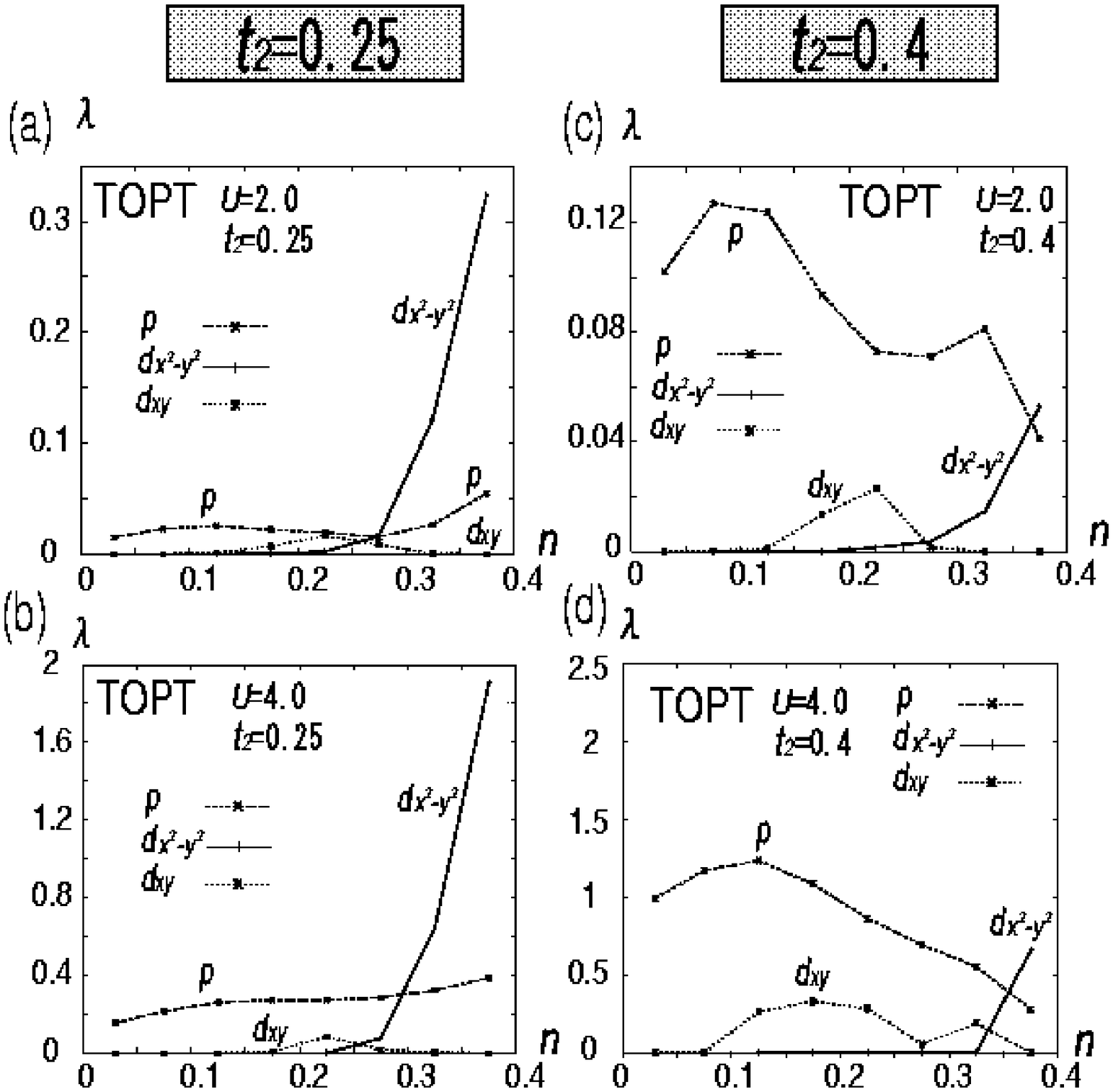,height=17cm}
\caption{The $U$-dependence of the eigenvalue $\lambda$ obtained from all terms (TOPT) within the third-order perturbation theory for the $p$-, $d_{x^2-y^2}$- and $d_{xy}$-wave pairing states. The hopping integral is fixed as $t_2$=0.25 and 0.4. A values of $U$ are 2.0 and 4.0. The temperature $T$ equals 0.008. The region of the density $n$ is from 0.03 to 0.375.}
\label{fig:8}
\end{figure}
\clearpage
\begin{figure}
\epsfile{file=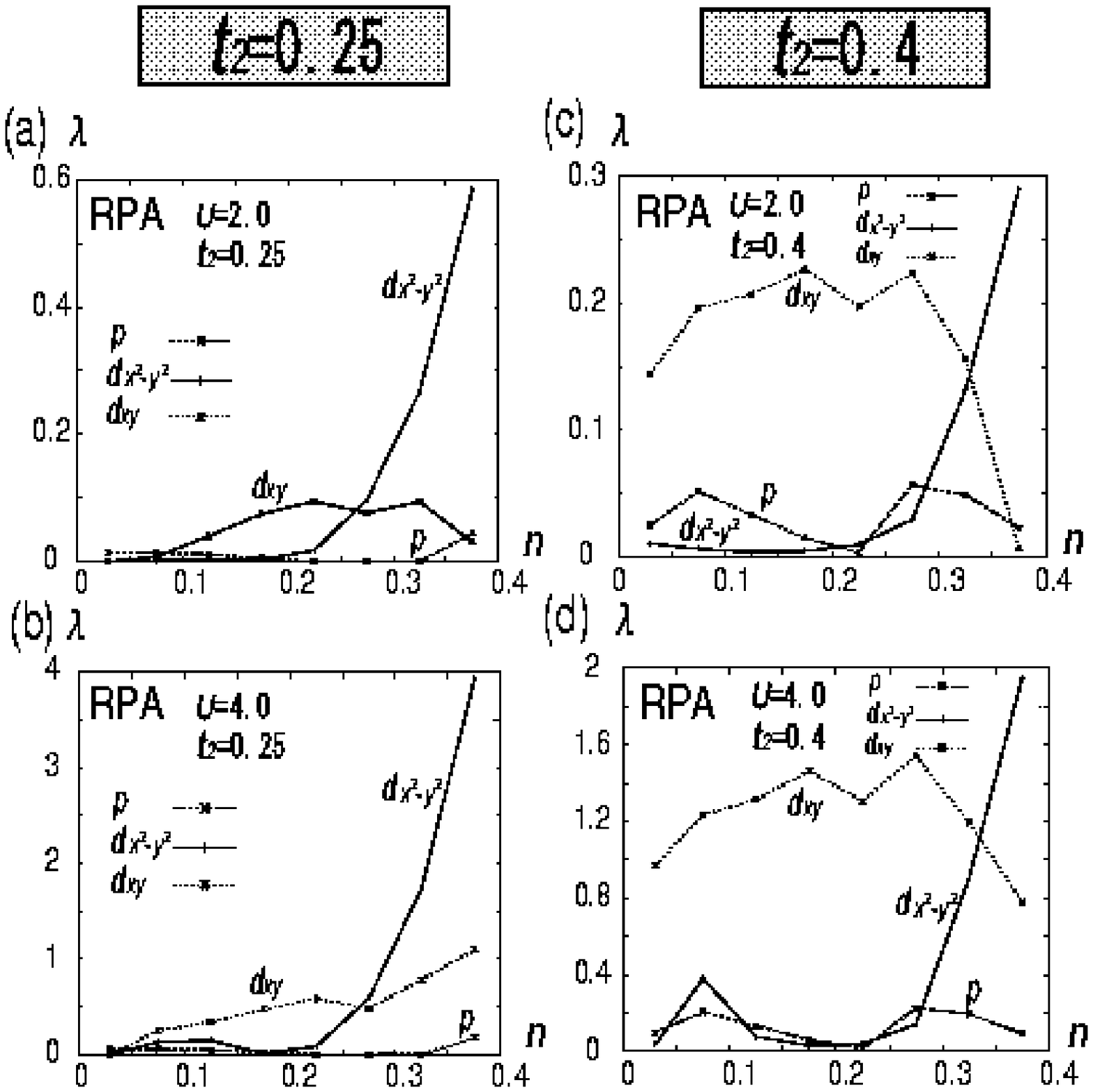,height=17cm}
\caption{The $U$-dependence of the eigenvalue $\lambda$ obtained from the RPA-like term for the $p$-, $d_{x^2-y^2}$- and $d_{xy}$-wave pairing states. The hopping integrals are fixed as $t_2$=0.25 and 0.4. The values of $U$ are 2.0 and 4.0. $T$ equals 0.008. The region of the density $n$ is from 0.03 to 0.375.}
\label{fig:9}
\end{figure}
\begin{figure}
\epsfile{file=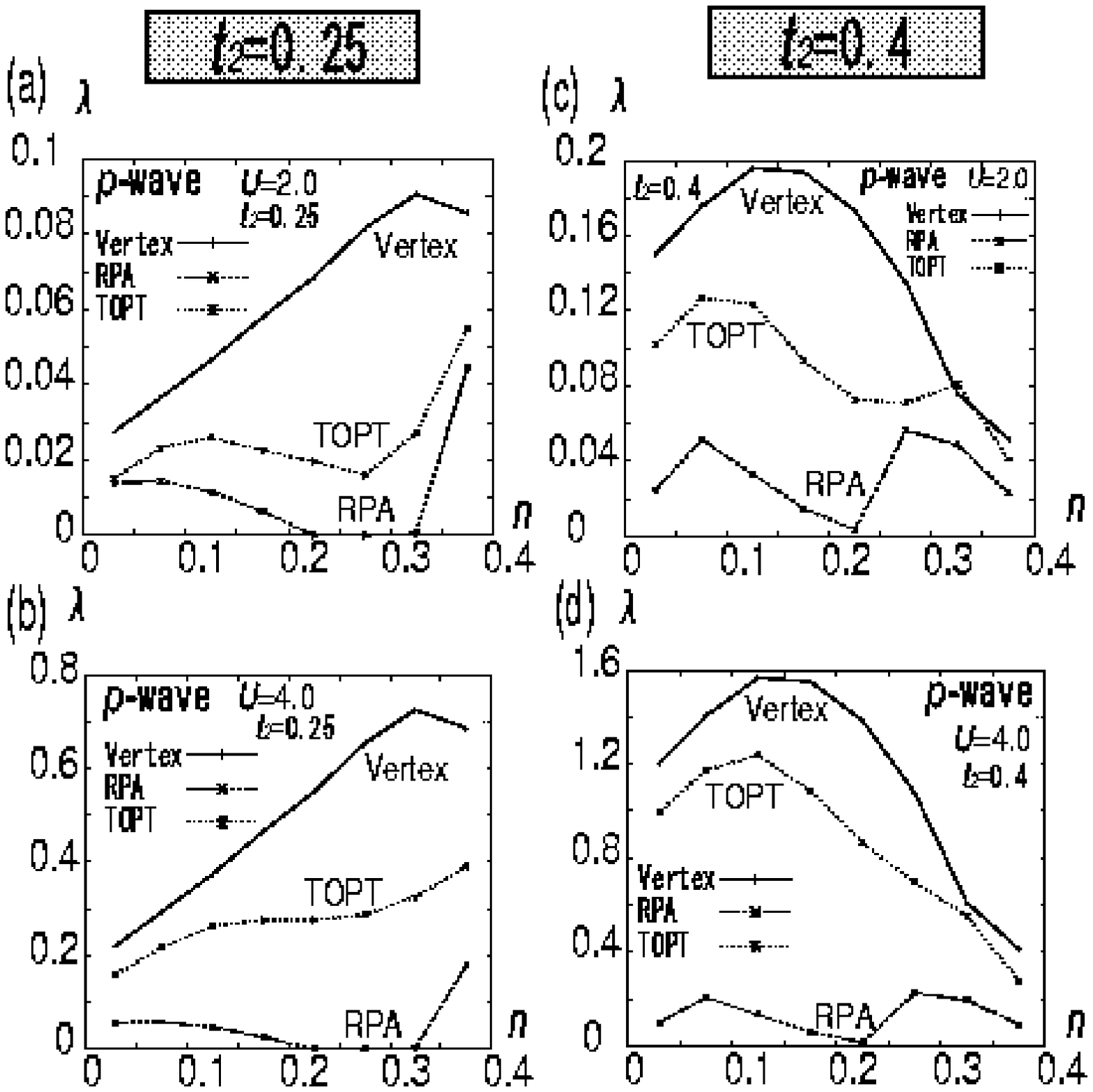,height=17.5cm}
\caption{The $U$-dependence of the eigenvalue $\lambda$ obtained from the all (TOPT), vertex and RPA-like terms for the $p$-wave pairing state. The hopping integral is fixed as $t_2$=0.25 and 0.4. The values of $U$ are 2.0 and 4.0. $T$ equals 0.008. The region of the density $n$ is from 0.03 to 0.375.}
\label{fig:10}
\end{figure}
\clearpage 
\begin{figure}
\epsfile{file=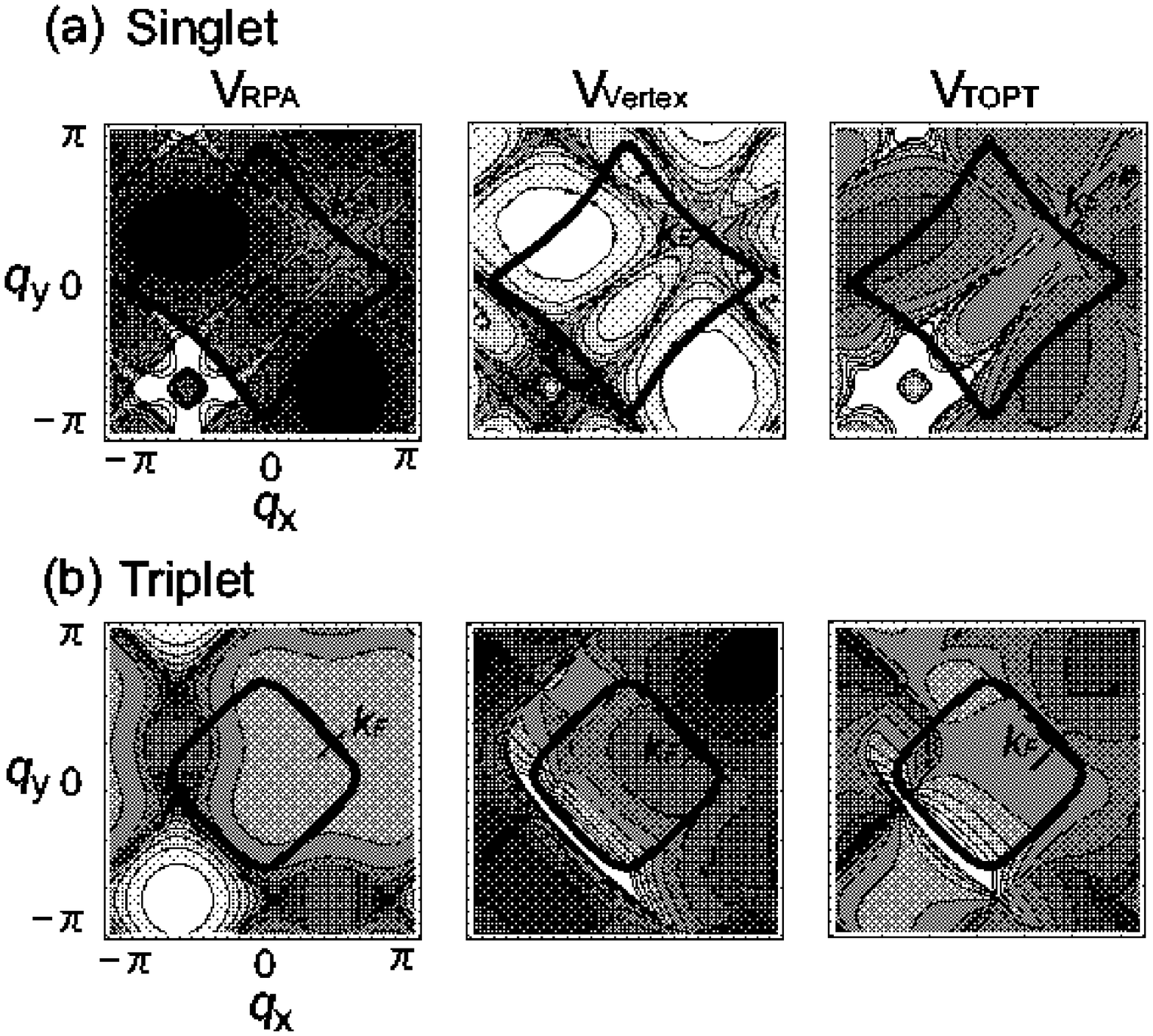,height=14.5cm}
\caption{The dependence of the singlet- and triplet-effective interaction $V$ on the wave number $q$ for all the terms (TOPT) $V_{\rm TOPT}$, the RPA-like term $V_{\rm RPA}$ and the vertex term $V_{\rm Vertex}$ on the momentum space $q$. The point $k=k_{\rm F}$ fixed on the Fermi surface and $q$ are an initial and a final state of the scattering. A lighter color shows a region of the stronger effective interaction. (a) The singlet effective interaction in the $d_{x^2-y^2}$-wave pairing state. The parameter sets are $T$=0.008, $U$=6.4, $t_2$=0.25 and $n$=0.225. (b) The triplet effective interaction in the $p$-wave pairing state. The parameter sets are $T$=0.008, $U$=6.4, $t_2$=0.25 and $n$=0.375.}
\label{fig:11}
\end{figure}
\end{document}